\documentclass{nature}

\usepackage[labelfont=bf]{caption}
\usepackage{graphicx}
\makeatletter
\let\saved@includegraphics\includegraphics
\AtBeginDocument{\let\includegraphics\saved@includegraphics}
\renewenvironment{figure}{\@float{figure}}
{\end@float}
\makeatother
\usepackage{amsmath}
\usepackage{comment}

\title{Tunable Giant Exchange Bias in an Intercalated Transition Metal Dichalcogenide}

\author{Spencer Doyle$^{1,2, \dagger}$, Caolan John$^{1,2, \dagger}$ Eran Maniv$^{1,2, \dagger}$, Ryan A. Murphy$^{3}$, Ariel Maniv$^{4,5}$, Sanath K. Ramakrishna$^{5}$, Yun-Long Tang$^{1,2,6}$, Ramamoorthy Ramesh$^{1,2,6}$, Jeffrey R. Long$^{2,3,6,7}$, Arneil P. Reyes$^{5}$ and James G. Analytis$^{1,2,*}$}

\begin{document}

\maketitle

\begin{affiliations}

% Affiliation 1 (UC Berkeley)
 \item Department of Physics, University of California, Berkeley, California 94720, USA
 
% Affiliation 2 (LBNL)
 \item Materials Science Division, Lawrence Berkeley National Laboratory, Berkeley, California 94720, USA
 
% Affiliation 3 (UC Berkeley)
 \item Department of Chemistry, University of California, Berkeley, California 94720, USA 
 
% Affiliation 4 (NRCN)
 \item NRCN, P.O. Box 9001, Beer Sheva, 84190, Israel
 
% Affiliation 5 (MagLab)
 \item National High Magnetic Field Laboratory, Tallahassee, Florida 32310, USA
 
% Affiliation 6 (UC Berkeley)
\item Department of Materials Science and Engineering, University of California, Berkeley, California 94720, USA
 
% Affiliation 7 (UC Berkeley)
\item Department of Chemical and Biomolecular Engineering, University of California, Berkeley, California 94720, USA

\end{affiliations}

\begin{abstract}
The interplay of symmetry and quenched disorder leads to some of the most fundamentally interesting and technologically important properties of correlated materials. It also poses the most vexing of theoretical challenges. Nowhere is this more apparent than in the study of spin glasses. A spin glass is characterized by an ergodic landscape of states - an innumerable number of possibilities that are only weakly distinguished energetically, if at all\cite{fischer1993spin,mydosh2014spin}. We show in the material Fe${_x}$NbS$_2$, this landscape of states can be biased by coexisitng antiferromagnetic order. This process leads to a phenomenon of broad technological importance: giant, tunable exchange bias. We observe exchange biases that exceed those of conventional materials by more than two orders of magnitude\cite{Nogues1999,nogues2005exchange}. This work illustrates a novel route to giant exchange bias by leveraging the interplay of frustration and disorder in exotic materials.  
\end{abstract}

Technologies that leverage correlated properties of quantum materials are one of the most active areas of research at the boundary of physics and engineering. One such property is exchange bias. It is a critical component to a variety of devices such as spin-valves, used extensively in high density magnetic storage\cite{kools1996exchange}, and has potentially more exotic applications, such as voltage-mediated magnetic switching for logic devices\cite{he2010robust}. Exchange bias manifests itself as a shift in the hysteresis loop of a magnetic system when cooled under an applied external field\cite{meiklejohn1956new}, and is observed in a diversity of systems, which, of particular importance to this study, include spin glasses\cite{Giri2011,Barnsley2013,Hudl2016}. Spin glasses are a phase of matter occurring in many strongly correlated systems, but differ from ordered ferro- or antiferromagnets in that their ground state is metastable, being one of many nearly degenerate states\cite{fischer1993spin,mydosh2014spin}. Central to these systems is frustration, which emerges as a result of site disorder\cite{nagata1979low,binder1986spin} or local competition between exchange interactions\cite{dekker1989activated}. The frustration protects the ergodicity of the system, until the spin glass transition is reached, at which point one state is settled upon. Understanding the dynamics of this state and the process by which it is chosen, remains a major theoretical question in the statistical mechanics of solids.

Theoretical challenges notwithstanding, the frozen state of the spin glass depends on its history, in particular the applied field in which it was cooled. This is the origin of its exchange bias: the magnetism of the frozen state is biased by the correlations of the spin glass. Typically this is very small, of order $\sim 0.01$ T. We report a giant exchange bias exceeding 3 T in Fe-intercalated 2H-NbS\textsubscript{2} when cooled in a field of 7 T. This value is not saturated, hinting that this family may host a record-breaking exchange bias at higher fields.

Fe$_{x}$NbS$_{2}$ is part of a larger class of transition metal dichalcogenides showing a variety of unexpected magnetic phenomena\cite{ghimire2018large,togawa2012chiral,kousaka2016long,checkelsky2008anomalous}. This material has been reported to possess antiferromagnetic (AFM) hexagonal ordering along the c-axis below its N\'eel temperature\cite{van1971magnetic}. For intercalation values less than x = 1/3, spin glass-like behavior has been observed in magnetization and heat capacity measurements\cite{Doi1991,yamamura2004heat,tsuji1999heat}.
Magnetization versus temperature measurements performed along the c-axis on Fe\textsubscript{x}NbS\textsubscript{2} for various intercalation values (x = 0.33, 0.31, 0.30, 0.28) are shown in Figure~\ref{fig:SG}a. The intercalation values were determined by energy dispersive X-ray spectroscopy (see supplementary information). For x = 0.33, corresponding to the fully packed Fe\textsubscript{1/3}NbS\textsubscript{2} structure, a clear AFM transition is observed, with a N\'eel temperature of approximately 42 K, corroborating previous measurements\cite{parkin19803I,parkin19803II,friend1977electrical,yamamura2004heat}.
For intercalation values below x = 0.33, field cooled (FC) and zero field cooled (ZFC) curves begin to separate, indicating the presence of a frozen moment. Both the onset temperature of this separation and the magnitude of the moment grow as Fe is removed. In addition, the magnetization is observed to relax with time on removal of the applied field for x $<$ 0.33, a characteristic of glassy systems (Figure~\ref{fig:SG}b)\cite{mydosh2014spin}. At x = 0.33, no relaxation is observed, as expected from the ultrafast spin dynamics in an AFM. The net magnetization and the relaxation timescale both increase as Fe is removed. The presence of a frozen moment and magnetic relaxation are both signatures of spin glass behavior. In addition, coercive hysteresis loops accompanied by a giant exchange bias are observed to onset and grow precipitously in the same range of compositions, as shown in Figure~\ref{fig:SG}c.

A further signature of spin glass behavior is the presence of thermal memory. By waiting at various temperatures while {\it cooling} the system, the magnetization measured upon {\it heating} the system is noticeably suppressed at the waiting points (Figure~\ref{fig:SG}d illustrates this effect for x = 0.30). This memory originates from the system relaxing into lower energy metastable states during a pause at a given temperature\cite{vincent2007ageing,mydosh2014spin}. Upon returning to this temperature, the system is annealed to the same local minimum it relaxed from, illustrating that both time and temperature are key players in determining the energy landscape. 
From magnetization FC/ZFC curves, relaxation dynamics, memory effects, and heat capacity (see supplementary information), we conclude that the system enters a glassy state for intercalation values just below the fully packed 1/3 structure. 

\begin{figure}[t]
    \centering
    \includegraphics[width=1\textwidth]{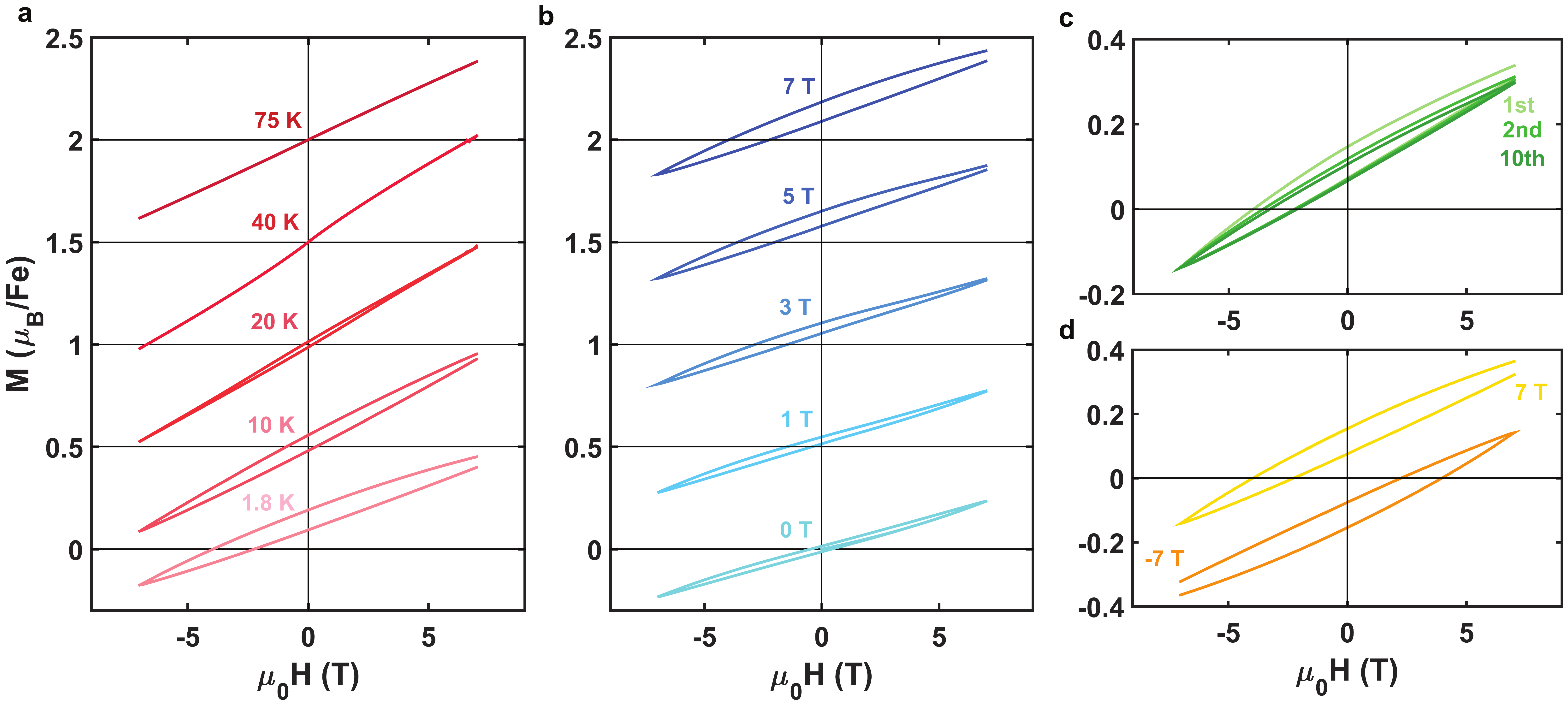}
    \caption{Magnetization versus magnetic field loops performed on the x=0.30 sample. All loops were taken from +7 T to -7 T back to +7 T. (a) Loops taken after field-cooling the sample in a 7 T field, and measuring at various temperatures. Each loop is offset on the y-axis by 0.5 $\mu_B$/Fe. The 1.8 K loop shows our largest observed exchange bias of 3.1 T. (b) Loops taken at 1.8 K after field-cooling the sample in various fields. (c) Demonstration of training in the sample. 10 consecutive hysteresis loops were taken, showing the slight decrease of the exchange bias and coercivity after the first loop. (d) Loops taken after field cooling in +7 T (top) and -7 T (bottom).}
    \label{fig:MvsH}
\end{figure}

Fe$_{0.30}$NbS$_{2}$ serves as an illustrative example of the exchange bias properties of these materials. Magnetic hysteresis loops are plotted in Figure~\ref{fig:MvsH}. Temperature dependent field sweeps were performed after field cooling in 7 T (Figure~\ref{fig:MvsH}a).
Below 20 K a significantly shifted hysteresis loop emerges due to the onset of an enhanced exchange bias. The bias, given by the average x-intercept, sharply increases to a value of 3.1 T at 1.8 K. To our knowledge, this is within 0.2 T of the largest reported exchange bias\cite{nayak2015design}. The exchange bias is proportional to the applied cooling field, as shown in Figure~\ref{fig:MvsH}b. 
It also shows a training effect such that the hysteresis loops shift on successive field cycles\cite{Nogues1999}, approaching a single location as a permanent bias field is obtained (Figure~\ref{fig:MvsH}c). Cooling in fields of $\pm$7 T results in no noticeable difference in exchange bias field, demonstrating that this is a symmetric effect (Figure~\ref{fig:MvsH}d). This symmetry indicates that the exchange bias is fully dependent on the cooling field, and thus has no spontaneous component\cite{migliorini2018spontaneous}.

\begin{figure}[t]
    \centering
    \includegraphics[width=0.8\textwidth]{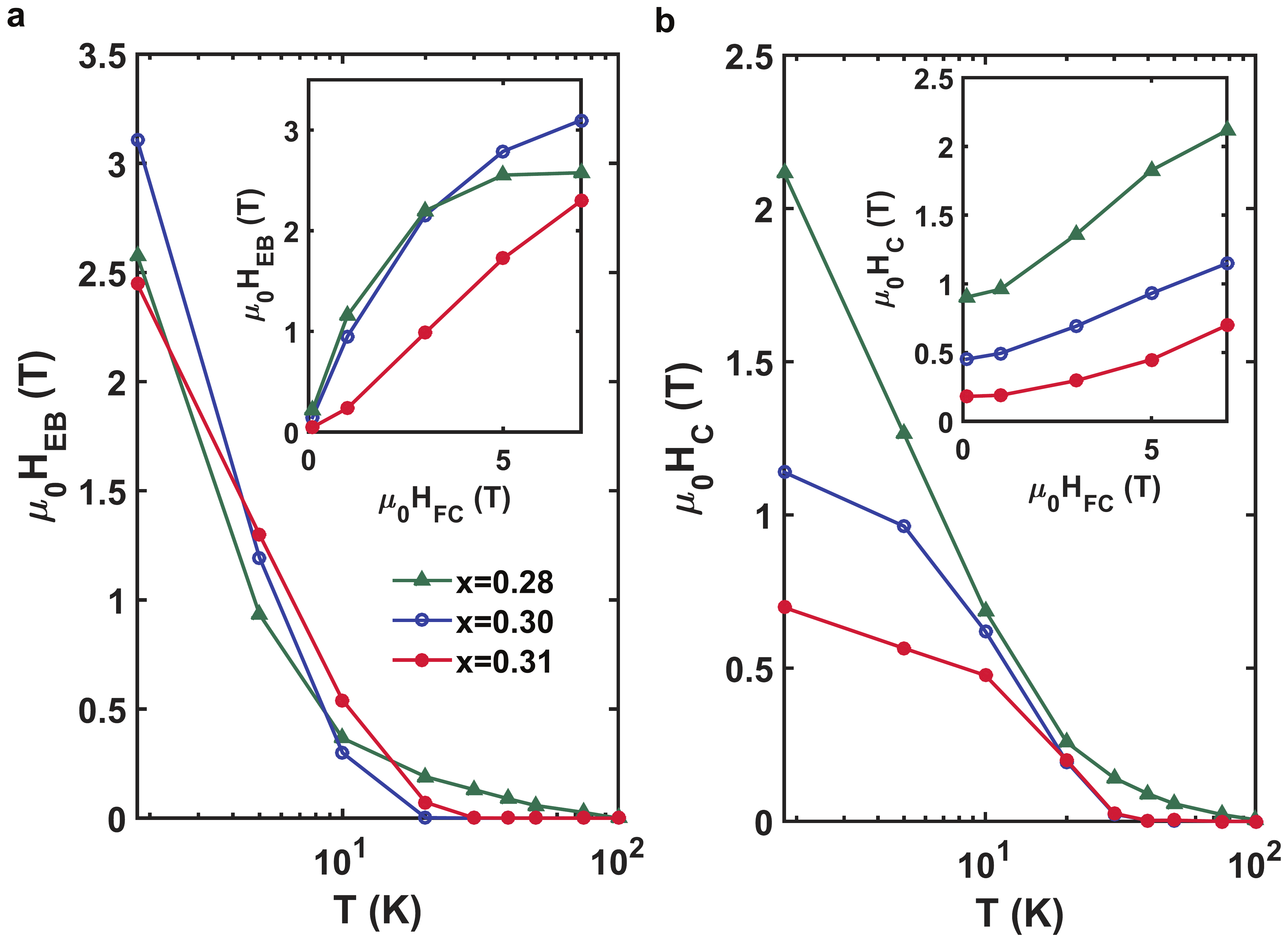}
    \caption{The intercalation tunability of Fe$_{x<1/3}$NbS$_{2}$. (a) Exchange bias as a function of temperature for samples x = 0.28, x = 0.30 and x = 0.31 (green triangles, blue empty circles, and red full circles respectively).  All scans were taken after field-cooling in 7 T. The inset shows the exchange bias field versus the cooling field. Note the lack of saturation in the x = 0.31 and x = 0.30 systems. (b) Coercive field as a function of temperature. All loops were taken after field-cooling in 7 T. The inset shows the coercive field as a function of cooled field. The increase in coercivity is a signature of an exchange bias system.}
    \label{fig:EB}
\end{figure}

The intercalation tunability of the exchange bias and coercivity is demonstrated in Figure~\ref{fig:EB}. Looking at the exchange bias field versus temperature (Figure~\ref{fig:EB}a), all three intercalation values show an increase in their exchange bias as the temperature decreases, with a notable upturn between 20-30 K. This upturn occurs well below the freezing point and at roughly the same temperature regime for all x values, suggesting that the mechanism responsible for the exchange bias does not exclusively stem from the glassy nature of the system. The inset shows the evolution of the bias with cooling field. For x = 0.28, the exchange bias appears fully saturated. For x = 0.30 and 0.31, however, the exchange bias is not saturated and we expect it to continue to increase if the samples were cooled in fields larger than 7 T. 
Figure~\ref{fig:EB}b illustrates the corresponding trends in coercivity. Each compound shows the onset of a large coercive field below the freezing temperature, with the lower intercalation values having a larger coercivity. This increase is correlated with the enhanced glassiness as Fe is removed from the pure AFM structure. Similarly, the coercive field in all samples increases with cooling field, an enhancement correlated with the appearance of the exchange bias\cite{Ali2007,Kosub2012}. Just as for the exchange bias, small variations in intercalation allow for control of the coercivity, with a difference in intercalation values of just $\Delta$x = 0.03 corresponding to an increase of nearly 1.5 T in the coercive field.

\begin{figure}[t]
    \centering
    \includegraphics[width=1\textwidth]{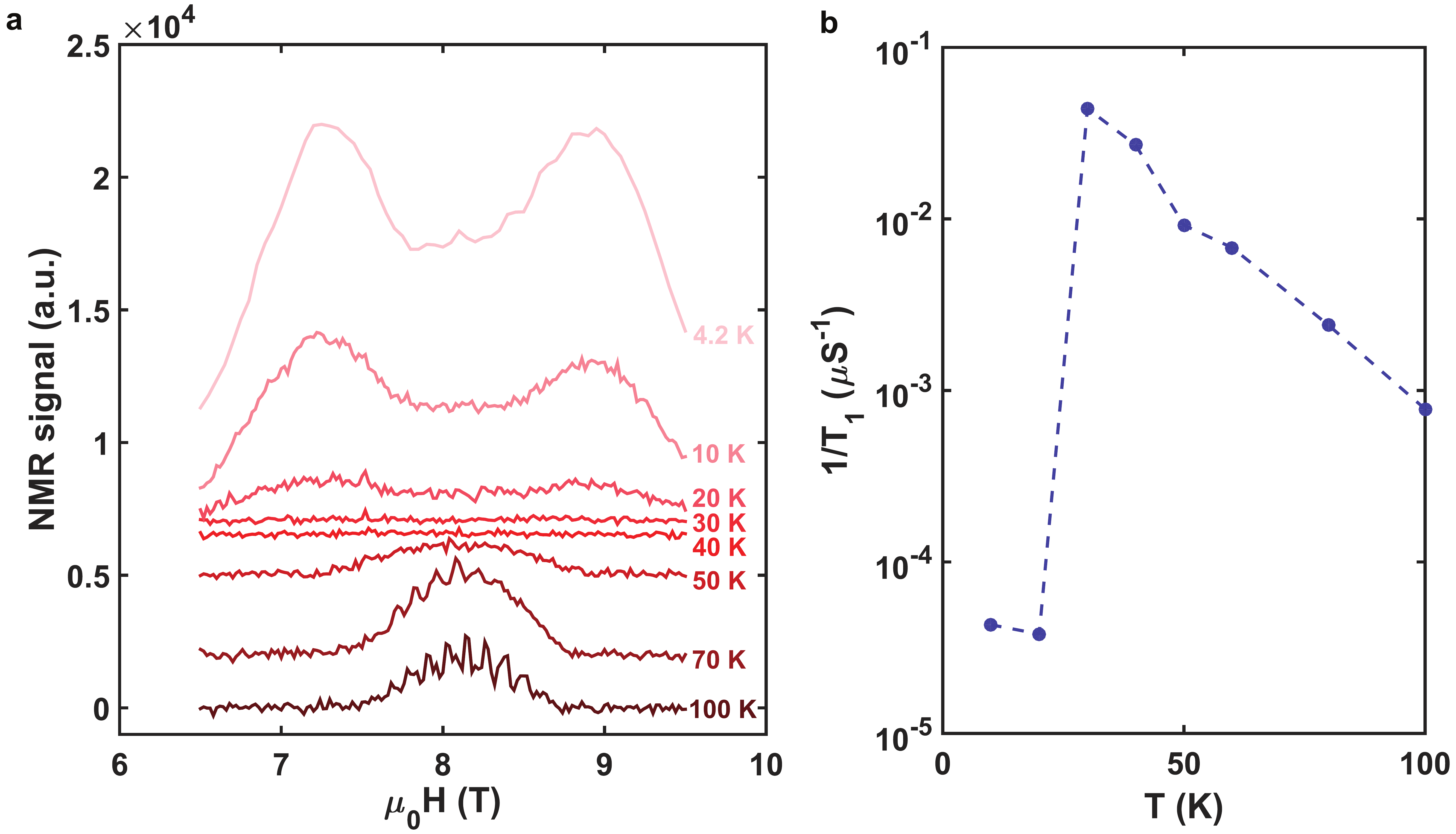}
    \caption{NMR measurements performed on the x = 0.30 sample. (a) Field-swept NMR spectra at 85 MHz for several temperatures from 100 K to 4.2 K. The magnetic field was applied along the c-axis of the sample. The nine expected $^{93}$Nb quadrupolar splittings are clearly resolved at 100 K, but broaden as the temperature is lowered. The quadrupolar coupling was extracted to be 1.25MHz. At temperatures below 25 K, two broad peaks indicative of a long-range AFM order emerge. (b) $1/T_1$ measured at 8.17T versus temperature, demonstrating a broad magnetic transition ending at around 30 K.}
    \label{fig:NMR}
\end{figure}

The majority of spin glass systems feature comparatively weaker bias fields~\cite{Ali2007,Giri2011,Barnsley2013,Hudl2016}. To better understand the underlying microscopic origin of the strong exchange bias, nuclear magnetic resonance (NMR) measurements were performed on a x = 0.30 sample.
NMR measurements have been previously performed on similar samples, but with different intercalation values\cite{okubo200793nb}.
Figure~\ref{fig:NMR}a shows the evolution of the $^{93}$Nb (I = 9/2, $\gamma$ = 10.405 MHz/T) field-swept spectra at 85 MHz as the sample is cooled. Above 100 K, the system is paramagnetic and the NMR spectrum shows a well-resolved quadrupolar splitting of 9 peaks. Owing to local magnetic disorder, these peaks broaden as the temperature is lowered, resulting in a featureless peak at 50 K. The broadening can be attributed to gradual local spin-freezing, suggestive of a spin-glass state. On further cooling, the signal intensity weakens until it is almost flat at around 40 K. This is due to extremely short relaxation times ($<100~\mu s$) near the magnetic transition temperature (Figure~\ref{fig:NMR}b). Below 20 K, the spectrum transforms into a double-peak structure, whose position is symmetric about the paramagnetic center. This is a signature of an AFM state, with the two peaks originating from the two sublattices where the hyperfine field ($\sim$1 T) adds to, and subtracts from, the applied magnetic field\cite{buttgen2012high}. Closer examination of the spectra below 20 K, reveals a broad background in between the two peaks. This compounded line-shape is consistent with what is expected from a spin density wave, an incommensurate AFM, or an AFM with a spin glass background. Though NMR spectra does not distinguish among these possibilities, the thermodynamic and magnetic properties of these materials strongly favor the latter--AFM and spin glass coexist. 

Additional information about the magnetic states in the system is provided by spin-lattice relaxation time ($T_1$) measurements, shown in Figure~\ref{fig:NMR}b. The jump in $1/T_1$ relaxation near 40 K is a canonical indicator of a magnetic phase transition, as electronic spin fluctuations, which couple to the nuclear spins, diverge near the phase transition. This is also noted in the temperature-dependent line shape change above and below 40 K, as seen in Figure~\ref{fig:NMR}a. In addition, the gradual increase in $1/T_1$ is correlated to the broadening of the NMR line shape below 100 K, as again, indicative of an increase in electronic fluctuations near the transition temperature. The gradual evolution into the multi-minima energy landscape of a spin glass leads to characteristically broad transitions\cite{fischer1993spin,mydosh2014spin}, thus the wide temperature region in which the fluctuations are significant provides local evidence of a glassy state. These two measurements provide supporting evidence of a glassy state in Fe$_{0.30}$NbS$_2$ at all temperatures below 40 K and an emergent long-range AFM order below 30 K. 

The combination of magnetization and NMR studies implies that a spin glass state and a long-range AFM order coexist. 
In all temperature dependent field sweeps (Figure~\ref{fig:EB}a) a large exchange bias emerges below 20 K, corresponding to the temperature where significant long-range AFM order is apparent in NMR. This is strong evidence that this coexistence is responsible for the giant exchange bias. This may be understood in the following way. The Sherrington-Kirkpatrick model of spin glasses successfully capture the ergodicity of the spin glass state, namely that each possible state in the landscape of possibilities is roughly interchangeable when looked at through the lens of spatial spin fluctuations. Above the spin glass transition, the possible states are roughly equivalent. Below the spin glass transition, this ergodicity is broken and only one state is chosen\cite{parisi1983order}. However, the other states are only weakly distinguished in energy, so that effects like exchange bias are generally small ($\sim 0.01$T). In contrast, in an AFM only one of two degenerate states is possible for a local spin, corresponding to distinct spin orientations. Our data suggests that this biases the glassy landscape, strongly distinguishing them in energy by leveraging the broken symmetry of the AFM, greatly enhancing the exchange bias. This can be seen directly in the x = 0.28 material, where a small exchange bias persists above this regime because the spin glass behavior onsets at much higher temperatures (see Figure~\ref{fig:EB} and Supplementary Materials). However, a significant enhancement of the exchange bias is only seen below 30K, the regime where AFM correlations are present.

Another remarkable feature of the Fe$_{x}$NbS$_2$ system is the extraordinary sensitivity to the intercalation value. With a $\Delta$x = 0.02 from the filled Fe$_{1/3}$NbS$_2$ structure, we see a shift from no exchange bias to values among the highest ever reported. Further intercalation changes from x = 0.31 to x = 0.30 results in a coercivity enhancement of 0.5 T. At the core of this sensitivity is the interplay between coexisting AFM and spin glass states. Fe$_{x}$NbS$_{2}$ is a candidate for low temperature applications, presenting a highly robust and tunable exchange bias induced even by very small cooling fields.

Exchange bias driven by the interplay of AFM and ergodicity is distinct from the canonical ferromagnetic/AFM bilayer picture. The present system does not rely on pinning at just an interface, but throughout the entire volume of the sample. Previous work has observed an enhanced bias field in a heterostructure of a ferromagnet and a spin glass\cite{Ali2007}. Our study demonstrates exchange bias without any ferromagnetic materials, appealing to a conceptually general mechanism; the cooperative action of spin glass and AFM order that compromises the ergodic landscape of the spin glass, but amplifies its unique properties. This observation of a giant, tunable exchange bias suggests a promising path to the discovery of larger bias field systems. Intercalated transition metal dichalcogenides present a vast playground to search for this physics.
In addition, this study opens the door for device applications of spin glasses, systems often considered to be of theoretical interest but with near absent applicability.

\newpage

\begin{figure}[t]
    \centering
    \includegraphics[width=1\textwidth]{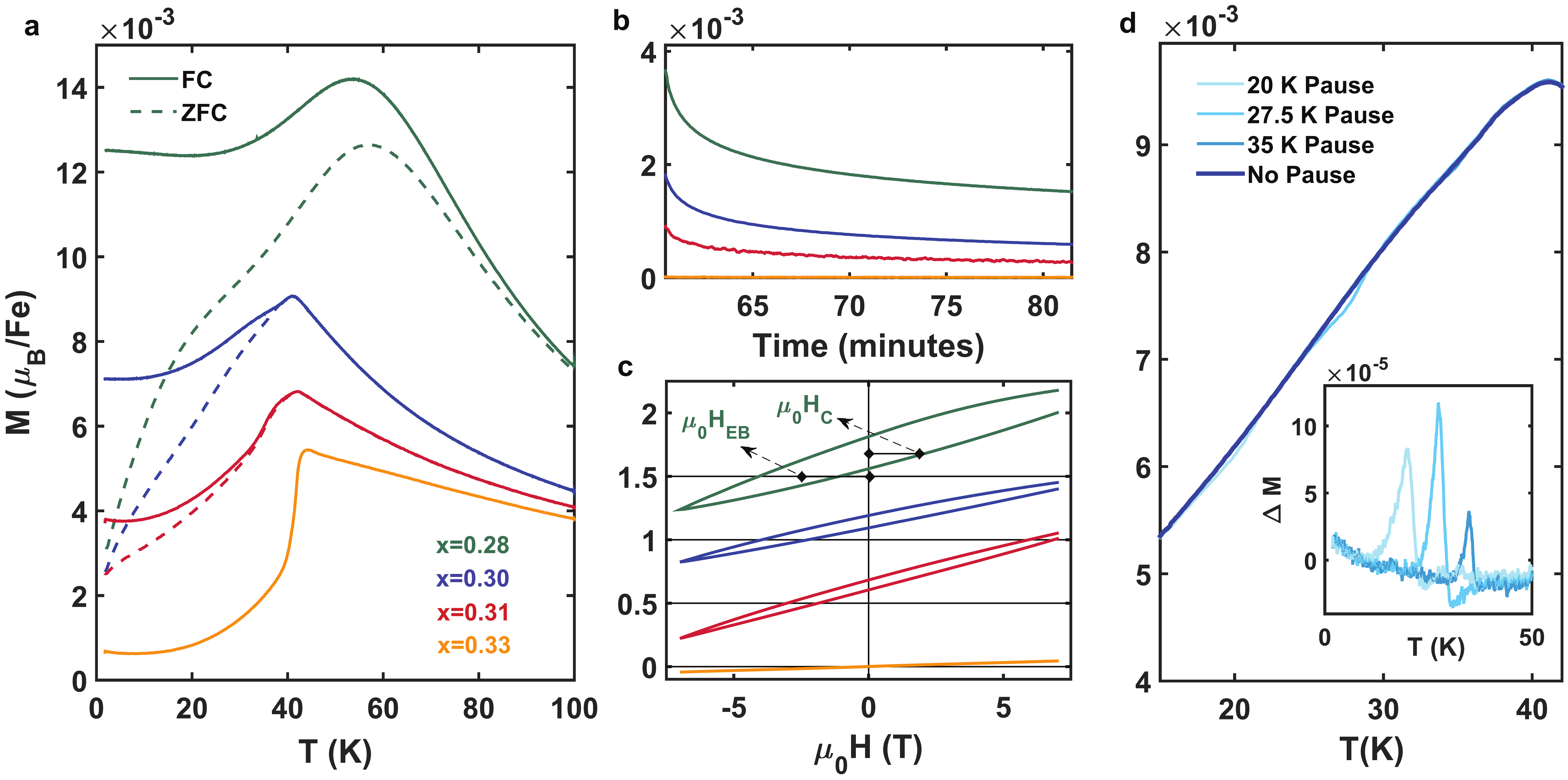}
    \caption{Magnetic characterization of Fe$_x$NbS$_2$ for x = 0.28, x = 0.30, x = 0.31, and x = 0.33. (a) 1000 Oe Magnetization versus Temperature curves for each intercalation value: both the FC (solid line) and ZFC (dashed line) curves are shown. The divergence of the FC and ZFC curves demonstrates the onset of glassy behavior. (b) Relaxation of the magnetization in all samples at 5 K after a 1 T field was applied for 1 hour. (c) Magnetization versus magnetic field loops at 1.8 K, for all x values, taken after field-cooling in a 7 T field. Each loop is offset on the y-axis by 0.5 $\mu_B$/Fe. The exchange bias field ($\mu_0H_{EB}$) and coercive field ($\mu_0H_{C}$) are marked for example on the x=0.28 hysteresis loop. (d) 1-hour (no-field) pause measurements performed on the x=0.30 sample at temperatures 20 K, 27.5 K, and 35 K. The inset shows $\Delta$M=M\textsubscript{no pause}$-$M\textsubscript{pause}, emphasizing the difference between the no-pause and pause measurements.}
    \label{fig:SG}
\end{figure}

\begin{addendum}
\item [Methods] Single crystals of Fe$_{x}$NbS$_2$ were synthesized using a chemical vapor transport technique. A polycrystalline precursor was prepared from iron, niobium, and sulfur in the ratio $x:1:2$ (Fe:Nb:S). The resulting polycrystalline product was then placed in an evacuated quartz ampoule with iodine as a transport agent (2.2 mg/cm$^3$), and put in the hot end of a two zone MTI furnace with temperature set points of 800 and 950 for a period of 7 days. High quality hexagonal crystals with diameters of several millimeters were obtained. 
Magnetization measurements were performed using a Quantum Design MPMS-3 system with a maximum applied magnetic field of 7 T.
Heat Capacity measurements were performed using a XENSOR AC-sensor in a Cryogen-free magnet system. 
Powder X-ray diffraction measurements were performed using a Rigaku Ultima-4 system with a Cu K-$\alpha$ radiation. 
The intercalation values were determined by energy dispersive X-ray spectroscopy using an Oxford Instruments X-MaxN 50 $mm^{2}$ system. 
NMR measurements were performed using the spin-echo technique, in the Condensed Matter NMR lab, at NHMFL, using a home-built NMR spectrometer with quadrature detection. The magnetic field was varied between 6 T and 11 T at various temperatures from 4.2 K to 150 K. Nuclear relaxation rates were measured by monitoring the growth of the spin-echo signal after introducing multiple saturating pulses (typically 200 ns), with variable delay and fitting the resulting curve to a single exponential.
\item This work was supported as part of the Center for Novel Pathways to Quantum Coherence in Materials, an Energy Frontier Research Center funded by the U.S. Department of Energy, Office of Science, Basic Energy Sciences. J.G.A. and E.M. were supported by the Gordon and Betty Moore Foundation’s EPiQS Initiative through Grant No. GBMF4374.
R.A.M. and J.R.L. were supported by the National Science Foundation through Award No. DMR-1611525.
A portion of this work was performed at the National High Magnetic Field Laboratory, which is supported by the National Science Foundation Cooperative Agreement No. DMR-1644779 and the State of Florida.
Microdiffraction measurements were done with the assistance of Camelia Stan in the Advanced Light Source beamline 12.3.2 which is a DOE Office of Science User Facility under contract no. DE-AC02-05CH11231.
\item [Author Contribution] S.D., C.J. and E.M. equally contributed to this work. S.D. and C.J. performed crystal synthesis and magnetization measurements. E.M. performed heat capacity and EDS measurements. Y.L.T. performed transmission electron microscopy measurements. A.M., S.K.R. and A.P.R. performed NMR measurements. S.D., C.J. and E.M. performed data analysis and wrote the manuscript with input from all coauthors.  
\item[Competing Interests] The authors declare no competing financial interests.
\item[Correspondence] Correspondence and requests for materials
should be addressed to J.G.A.~(email: analytis@berkeley.edu).
\end{addendum}

\bibliographystyle{naturemag}

\end{document}